\documentclass[epj]{webofc}
\usepackage[varg]{txfonts}   

\usepackage{bm}
\usepackage{amssymb}
\usepackage{amsmath}
\usepackage{enumerate}
\usepackage{epsf}
\usepackage[usenames]{color}

\wocname{EPJ Web of Conferences}
\def\S{\mathcal{S}}

\def\eRM{{\mathrm e}}
\def\dRM{{\mathrm d}}

\def\eps{\varepsilon}

\allowdisplaybreaks
\allowbreak
\newcommand{\fp}[1]{FP$ {\textrm{#1}}$}

\begin{document}
\selectlanguage{english}
\title{Advection of a passive scalar field by turbulent compressible fluid: renormalization group analysis near $d=4$}%
%

\author{N. V. Antonov\inst{1}\fnsep\thanks{\email{n.antonov@spbu.ru}} \and
        N. M. Gulitskiy\inst{1}\fnsep\thanks{\email{n.gulitskiy@spbu.ru}} \and
        M. M. Kostenko\inst{1}\fnsep\thanks{\email{kontramot@mail.ru}} \and
        T. Lu\v{c}ivjansk\'y\inst{2}\fnsep\thanks{\email{tomas.lucivjansky@uni-due.de}}
        }

\institute{Department of Theoretical Physics, Faculty of Physics,  
Saint-Petersburg State University, 7/9~Universitetskaya nab., St. Petersburg, 199034 Russia
\and
Faculty of Sciences, P.J. \v{S}af\'arik
University, Moyzesova 16, 040 01 Ko\v{s}ice, Slovakia,\\ and
Fakult\"at f\"ur Physik, Universit\"at Duisburg-Essen, D-47048 Duisburg, Germany
}

\abstract{
The field theoretic renormalization group (RG) and the operator product
expansion (OPE) are applied to the model of a density field advected by a random turbulent
velocity field. The latter is governed by the stochastic Navier-Stokes equation
for a compressible fluid. 
The model is considered near the special space dimension $d=4$. 
It is shown that various correlation functions of the scalar field
exhibit anomalous scaling behaviour in the inertial-convective range.
The scaling properties in the RG+OPE approach are related to fixed points of the renormalization group equations. 
In comparison with physically interesting case $d=3$, at $d=4$ additional Green function has divergences which affect the existence and stability of fixed points. 
From calculations it follows that a new regime arises there and then by continuity moves into $d=3$.
The corresponding anomalous exponents are identified with scaling
dimensions of certain composite fields and can
be systematically calculated as series in $y$ (the exponent, connected with random force) and $\varepsilon=4-d$.
All calculations are performed in the leading one-loop approximation.
}

\maketitle
{\section{Introduction}\label{sec:intro}}
Understanding of fully developed turbulence is a very complex problem. The phenomenon is characterized by the cascade of energy from large scales 
toward smaller ones, where viscosity plays an important role and the dissipation effects dominate, and intermittency, which 
leads to anomalous scaling, i.e., singular behaviour of various statistical quantities as functions of the integral turbulence scales~\cite{Legacy}.
A very powerful technique, which is applicable to different models of critical phenomena, characterized by scaling behaviour, is 
field theoretic renormalization group (RG) and operator product expansion (OPE); see the monographs~\cite{Vasiliev,Zinn,turbo,Tauber}.
In the RG+OPE scenario, anomalous
scaling emerges as a consequence of the existence in the model some composite fields (``composite operators'' in the
quantum-field terminology) with negative scaling dimensions~\cite{JphysA}. 
In a number of papers the RG+OPE approach was applied to the case of passive advection by 
Kraichnan's ensemble (velocity field is taken isotropic, Gaussian, not correlated in time,
having a power-like correlation function, and the fluid was assumed to be incompressible), see~\cite{Kraich1,GK,RG,JJ15},
and by numerous of its generalizations: large-scale anisotropy, helicity,
compressibility, finite correlation time, non-Gaussianity, and
more general form of nonlinearity~\cite{JphysA,V96-mod,amodel,AG12,J13,Uni3,VectorN}.
Also this approach can be generalized to the case of non-Gaussian velocity field, governed by the stochastic Navier-Stokes equation~-- both for the scaling behaviour of the velocity field itself and for the advection of passive fields by this ensemble~\cite{NSpass,Ant04,AK14,AK15,AGM,JJR16}.

Up to present day a majority of works on fully developed turbulence are concerned with an incompressible fluid. 
However, some treatment has been applied to the compressible case too~\cite{VM,tracer2,tracer3,AK2,LM,St,MTY}. 
All of these works indicates a large influence of compressibility both on velocity field itself and passively advected quantities. Furthermore, they show us a necessity of further investigations focused on the compressibility.

In this paper an application of the field
theoretic renormalization group onto the scaling regimes of a compressible fluid, whose
behaviour is governed by a proper generalization of the stochastic Navier Stokes equation~\cite{ANU97}, is presented. 
Following~\cite{HN96}, we employ a double expansion scheme. Here the formal expansion 
parameters are $y$, which describes the
scaling behaviour of a random force, and $\eps=4-d$, i.e., a deviation from the space dimension $d=4$.
\newline

{\section{Description of the model and field theoretic formulation} \label{sec:desc}}
The Navier-Stokes equation for a viscid compressible fluid can be written in the following form:
\begin{equation}
\rho\nabla_{t} v_{i} = \nu_{0} [\delta_{ik}\partial^{2}-
\partial_{i}\partial_{k}] v_{k}
+ \mu_0 \partial_{i}\partial_{k} v_{k} - \partial_{i} p + \eta_{i},
\label{NS}
\end{equation}
where
\begin{equation}
\nabla_{t} = \partial_{t} + v_{k} \partial_{k}
\label{nabla}
\end{equation}
is the Lagrangian derivative, 
$\rho$ is the fluid density, $v_i$ is an $i$-th component of the velocity field $\bm v$,
$\partial_{t} =
\partial /\partial t$, $\partial_{i} = \partial /\partial x_{i}$, 
$\partial^{2} =\partial_{i}\partial_{i}$ is the Laplace operator, $p$ is the pressure field,  and $\eta_i$ is the density of an external force. 
The fields $v_i$, $ \eta_i$, $\rho$, and $p$ depend on $x=\{t,{\bm x}\}$ with
${\bm x}=\{x_1,x_2,\dots x_d\}$, $i=1\dots d$, where $d$ is the dimensionality of space.
The constants $\nu_{0}$ and $\mu_{0}$ are two
independent molecular viscosity coefficients; see\cite{LL}. Summations over repeated vector indices are always implied.

The model must be augmented by two additional equations, namely a continuity equation and an equation of state between 
deviations $\delta p$ and $\delta \rho$ from the equilibrium values~\cite{AK14,VN96}. Introducing the scalar field $\phi = c_0^2 \ln (\rho/\overline{\rho})$,
where parameter $c_0$ is an adiabatic speed of sound and $\overline{\rho}$ denotes the mean value of $\rho$, the model can be rewritten
in terms of two coupled equations:
\begin{align}
\nabla_{t} v_{i} &=
\nu_{0} [\delta_{ik}\partial^{2}-\partial_{i}\partial_{k}]
v_{k}\! +\! \mu_0 \partial_{i}\partial_{k} v_{k} -\!
\partial_{i} \phi\! +\! f_{i},
\label{ANU} \\
\nabla_{t} \phi &= -c_{0}^{2} \partial_{i}v_{i}.
\label{ANU1}
\end{align}

The turbulence is modeled by an external force -- it is assumed to be a random variable, which
mimics the input of the energy into the system on the large scale $L$. Its precise
form is believed to be unimportant and is usually considered to be a random Gaussian variable with zero mean
and the correlator
\begin{equation}
  \langle f_i(t,{\bm x}) f_j(t',{\bm x}') = \frac{\delta(t-t')}{(2\pi)^d} \int_{k>m} \dRM^d{\bm k}\mbox{ } \widetilde{D}_{ij}({\bm k})
  \eRM^{i{\bm k}\cdot({\bm x-\bm x'})},
\label{ff}
\end{equation}
where
\begin{equation}
  \widetilde{D}_{ij}({\bm k}) = g_{10} \nu_0^3 k^{4-d-y} \biggl\{  P_{ij}({\bm k}) + \alpha Q_{ij}({\bm k}) \biggl\}.
  \label{power}
\end{equation} 
Here $P_{ij}({\bm k}) = \delta_{ij}-k_ik_j/k^2$ and $Q_{ij}({\bm k})=k_ik_j/k^2$ are the transverse 
and longitudinal
projectors, $k=|{\bm k}|$, amplitude $\alpha$ is a free parameter,
an exponent $y$ plays a role of a formally small expansion parameter, and $g_{10}$ is a coupling 
constant; Dirac delta function ensures the Galilean invariance~\cite{turbo}. 

The expressions~\eqref{ANU}~-- \eqref{power} form a complete set of the equations needed to describe a compressible fluid.
The passive advection of a scalar density field $\theta(x)\equiv \theta(t,{\bm x})$ by this velocity field is described by the equation
\begin{equation}
\partial _t\theta+ \partial_{i}(v_{i}\theta)=\kappa _0 \partial^{2} \theta+f,
\label{density1}
\end{equation}
where $\kappa_0$ is the molecular diffusivity coefficient, ${\bm v}$ obeys the expressions~\eqref{ANU}~-- \eqref{power}, and $f = f(x)$ is a
Gaussian noise with zero mean and given covariance
\begin{equation}
\langle f(x)f(x') \rangle = \delta(t-t')\, C({\bm r}/L), \quad
{\bm r}= {\bm x} - {\bm x}'.
\label{noise}
\end{equation}
Here $C({\bm r}/L)$ is some function which is finite at $({\bm r}/L)\to 0$ and rapidly decaying as $({\bm r}/L)\to\infty$. 

According to the general theorem~\cite{Vasiliev,Tauber}, the stochastic problem~\eqref{ANU}~-- \eqref{noise} is equivalent to the field theoretic
model with De Dominicis-Janssen action functional for the doubled set of fields $\Phi=\left\{{\bm v}, {\bm v}',\phi, \phi',\theta,\theta'\right\}$:
\begin{equation}
{\cal S}(\Phi)= {\cal S}_{\theta}(\theta, \theta', {\bm v}) +
{\cal S}_{\bm v}({\bm v}, {\bm v}', \phi, \phi'),
\label{Fact}
\end{equation}
where
\begin{align}
{\cal S}_{\theta} (\theta, \theta', {\bm v}) = \frac{1}{2} \theta' D_{f} \theta' +
\theta' \left\{ - \partial_{t}\theta -
\partial_{i}(v_{i}\theta) +\kappa _0 \partial^{2} \theta \right\}
\label{Dact}
\end{align}
is the action functional for the stochastic problem
(\ref{density1}),
(\ref{noise}) at fixed ${\bm v}$. The second term in (\ref{Fact}) represents the velocity statistics:
\begin{align}
  \S_{\bm v}(\Phi) & = \frac{v_i' D_{ik} v_k'}{2} 
  +v_i' \biggl\{
  -\partial_t v_i - v_j\partial_j v_i + \nu_0[\delta_{ik}\partial^2 - \partial_i \partial_k]v_k
  +u_0 \nu_0 \partial_i \partial_k v_k - \partial_i \phi
  \biggl\} \nonumber \\  &
  +\phi'[-\partial_t \phi + v_j\partial_j \phi + v_0 \nu_0 \partial^2 \phi - c_0^2 (\partial_i v_i)].
  \label{eq:action} 
\end{align}
The function $D_f$ in~\eqref{Dact} is the correlation function~\eqref{noise}, the function $D_{ik}$ in~\eqref{eq:action} is given by~(\ref{power}).
Hereinafter integrals over the spatial variable 
${\bm x}$ and the time variable $t$, as well as summation over repeated indices, are implicitly assumed. 
Also we passed to the new dimensionless parameter
$u_0=\mu_0/\nu_0>0$ and introduced a new term
$v_0 \nu_{0} \phi' \partial^{2}\phi$ with the another positive
dimensionless parameter $v_0$, which is needed for the renormalization procedure. 
The standard methods of quantum field theory, such as Feynman diagrammatic technique and 
renormalization group procedure, can be applied to the action~(\ref{eq:action}).

The model is considered in the special space dimension $d=4$. Usually $d$ plays a passive role and the actual perturbative
parameter is $y$ [see~\eqref{power}]. But 
if $d=4$, the situation is similar to the analysis of the incompressible Navier-Stokes equation near space dimension $d=2$, see~\cite{HN96,AHKV03,AHH10}~-- 
an  additional divergence appears in the 1-irreducible Green function $\left\langle v'v'\right\rangle_{\text{1-ir}}$. 
To keep the model renormalizable at $d=4$, the kernel function $\widetilde{D}_{ij}({\bm k})$ in~(\ref{ff}) has to be replaced by
\begin{equation}  
  D_{ij}({\bm k})=g_{10} \nu_0^3 k^{4-d-y} \biggl\{
  P_{ij}({\bm k}) + \alpha Q_{ij}({\bm k})
  \biggl\} 
  + g_{20} \nu_0^3 \delta_{ij},
  \label{eq:correl2}
\end{equation}
and a regular expansion in both $y$ and $\eps=4-d$ should be constructed. 
Here $g_{20}$ is another coupling constant and the new term in the right hand side absorbs the divergent contributions from the function $\left\langle v'v'\right\rangle_{\text{1-ir}}$.

{\section{Feynman diagrammatic technique and renormalization of the model} \label{sec:technique}}

The perturbation theory of the model can be expressed in the standard Feynman diagrammatic expansion~\cite{Zinn,Vasiliev}.
Bare propagators are read off from the inverse matrix of the Gaussian (free) part of the action functional. The nonlinear part of the 
differential equation defines the interaction vertices.

The ultimate goal of the theory is to find the inertial range asymptotic behaviour of the measurable quantities, e.g., pair correlation 
functions of composite operators, built from the advected field $\theta$. In the RG+OPE approach they are connected with IR attractive
fixed points of RG equations. To investigate them 
we will follow the logic of~\cite{AK14,AK15} and~\cite{ANU97}: since the renormalization constants $Z_i$ and the RG functions of the 
reduced model~\eqref{eq:action} do not depend on the parameters, connected with the advection of passive field~\eqref{Dact}, it is possible
to divide our task in two parts. The first part is devoted to investigation of the stochastic Navier-Stokes equation itself, i.e., we will deal
only with the action functional~\eqref{eq:action}. And only after the establishing of the IR attractive fixed points for the action 
functional $\S_{\bm v}$ we will add to our consideration the action functional $\S_\theta$ [see~\eqref{Dact}] which is 
responsible for the advection of the density field.

\begin{figure}[t]
   \centering
   \begin{tabular}{c c}
     \includegraphics[width=6.1cm]{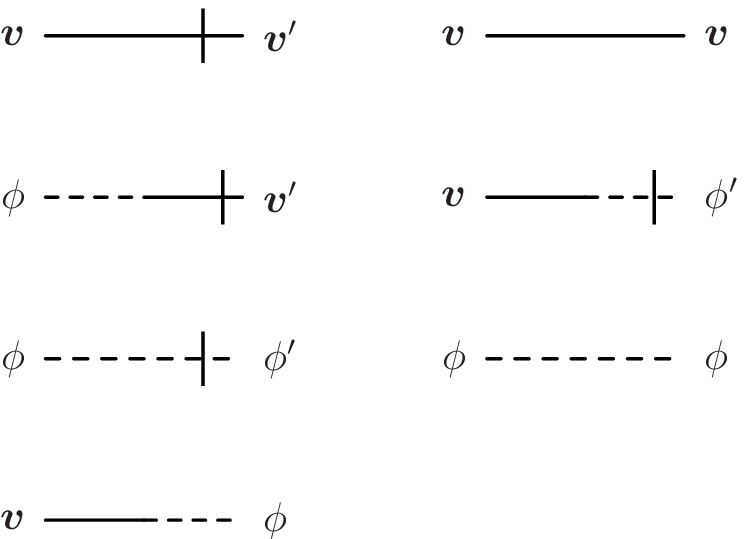}
     &
     \includegraphics[width=6.1cm]{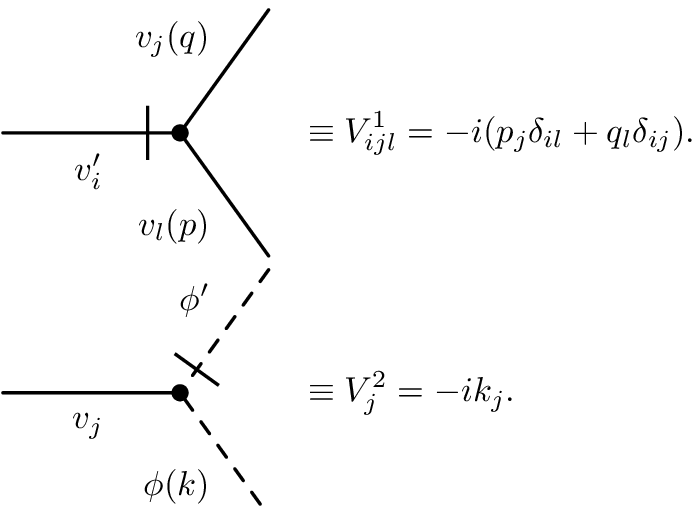}  
      \end{tabular}
   \caption{Graphical representation of the bare propagators and interaction vertices in the model~(\ref{eq:action})}
   \label{fig:prop_vertex}
\end{figure}

The graphical representation for the Feynman rules of the model ${\cal S}_{\bm v}$ is depicted in Fig.~\ref{fig:prop_vertex}. 
\noindent
In the frequency-momentum representation  the propagator functions attain the form:
\begin{eqnarray}
\langle vv' \rangle_{0} &=& \langle v'v \rangle_{0}^{*}= P({\bm k})
\epsilon_{1}^{-1} + Q({\bm k}) \epsilon_{3} R^{-1}, \quad
\langle vv \rangle_{0} =  P({\bm k}) \frac{d_1^{f}}{|\epsilon_{1}|^{2}}
+ Q({\bm k}) d_2^{f} \left|\frac{\epsilon_{3}}{R}\right|^{2},
\nonumber \\
\langle \phi v' \rangle_{0} &=& \langle v' \phi \rangle_{0}^{*}= -
\frac{{\rm i}c_{0}^{2}{\bm k}}{R}, \quad
\langle  v\phi' \rangle_{0} = \langle \phi'v \rangle_{0}^{*}=
\frac{{\rm i}{\bm k}}{R}, \quad 
\langle  v\phi \rangle_{0} = \langle  \phi v \rangle_{0}^{*} =
\frac{ {\rm i} c_{0}^{2} d_2^{f}\epsilon_{3} {\bm k}} {|R|^{2}} ,
\nonumber \\
\langle  \phi\phi' \rangle_{0} &=& \langle \phi'\phi \rangle_{0}^{*}=
\frac{\epsilon_{2}}{R}, \quad
\langle  \phi\phi \rangle_{0} = \frac {c_{0}^{4} k^{2}d_2^{f}}
{|R|^{2}}, \quad 
\langle  \phi'\phi' \rangle_{0} = \langle  v'\phi' \rangle_{0} =
\langle  v'v' \rangle_{0} = 0,
\label{lines}
\end{eqnarray}
where the vector indices of the fields and the projectors are omitted;
\begin{align}
\centering
\epsilon_{1}  &= -{\rm i}\omega +\nu_0 k^{2},& \quad 
\epsilon_{2}&=-{\rm i}\omega+ u_0\nu_0 k^{2},& \quad
d^{f}_1 &= g_{10}\nu_0^{3}\, k^{4-d-y}+g_{20}\nu_0^3,
\nonumber \\
\epsilon_{3}  &= -{\rm i}\omega+ v_0\nu_0 k^{2},& \quad 
R&=\epsilon_{2}\epsilon_{3}+c_{0}^{2}k^{2}, &\quad
d^{f}_2 &= \alpha [g_{10}\nu_0^{3}\, k^{4-d-y}]+g_{20}\nu_0^3.
\label{energies}
\end{align}
\pagebreak 

\begin{figure}[t]
   \centering
   \begin{tabular}{c c c}
   \raisebox{5.5ex}{\includegraphics[width=6.1cm]{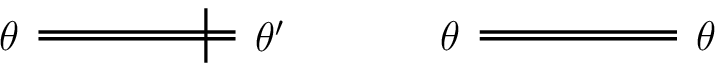}}  
  & \quad\quad &
   \includegraphics[width=3.5cm]{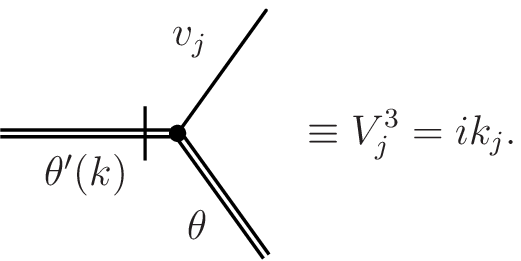}
    \end{tabular}
  \caption{Graphical representation of the bare propagators and interaction vertex in the model~(\ref{Dact})}
  \label{V3}
\end{figure}

Taking into account the second part of the action functional ${\cal S}_\theta$, one should add one vertex and two
propagator functions to the diagrams on Fig.~\ref{fig:prop_vertex}, see Fig.~\ref{V3}. In the frequency-momentum representation they read
\vspace{-.5em}
\begin{align}
\langle \theta \theta' \rangle _0=  \langle \theta' \theta \rangle_0^* =
\frac{1} {-{\rm i}\omega +\kappa _0 k^2},
\quad \quad
\langle \theta \theta \rangle _0 = \frac {C({\bm k}L)}
{\omega^{2} +\kappa_0^{2} k^4},
\label{lines3}
\end{align}
where $C({\bm k}L)$ is Fourier transform of the function $C({\bm r}/L)$, see~\eqref{noise}.

Hereinafter solid line denotes the field ${\bm v}$, solid line with a slash~-- the field ${\bm v'}$, dotted line~-- the
field $\phi$, dotted line with a slash~-- the field $\phi'$, double solid line~-- the field $ \theta $, double solid line with a slash~-- the field $ \theta'$.

\begin{figure}[b]
   \centering
   \begin{tabular}{c c c c}
   \includegraphics[width=2.5truecm]{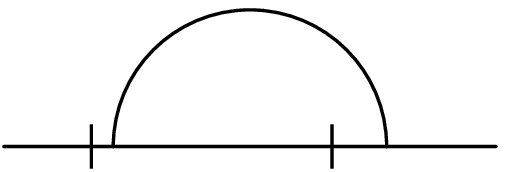}
   \quad &
   \includegraphics[width=2.5truecm]{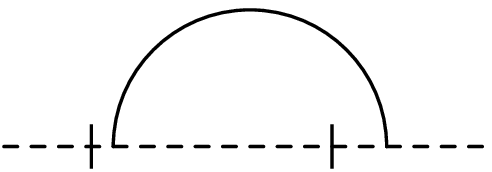}
   \quad &
   \includegraphics[width=2.3truecm]{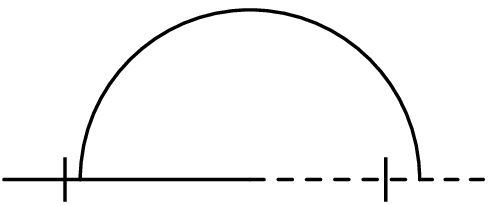}
   \quad &
   \includegraphics[width=2.7truecm]{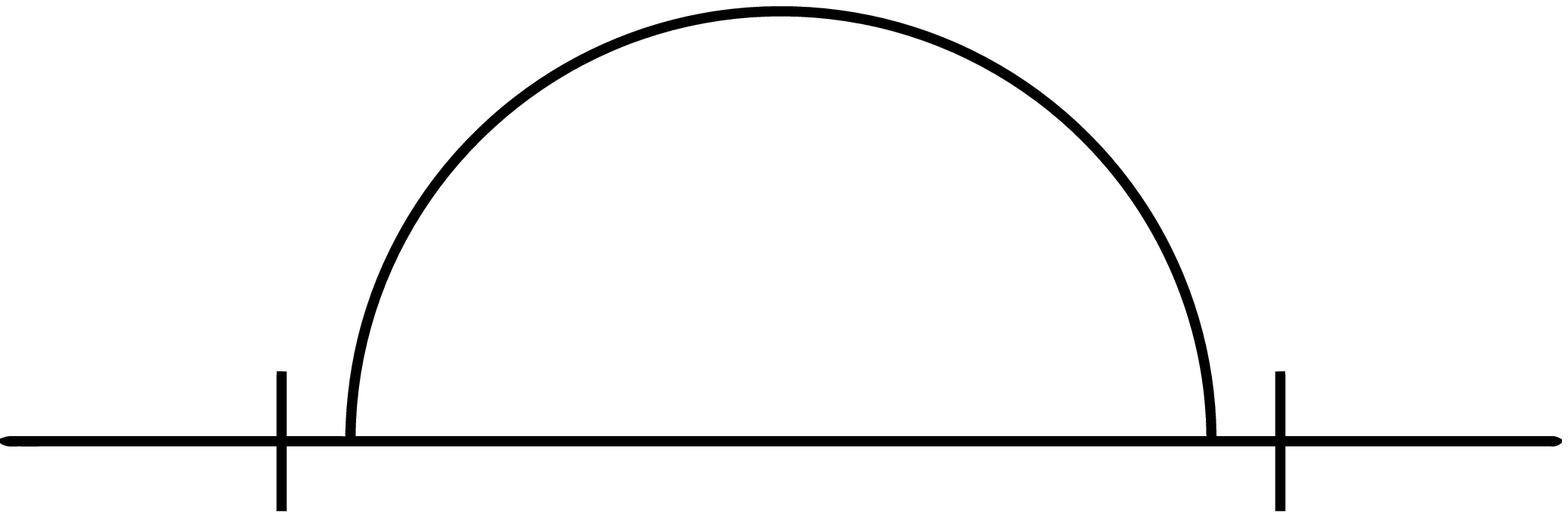} \\
   \includegraphics[width=2.3truecm]{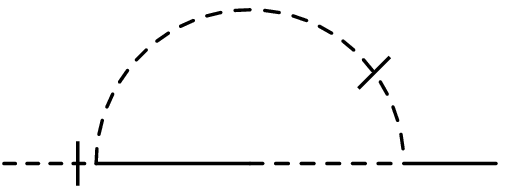}
   \quad &
   \includegraphics[width=2.3truecm]{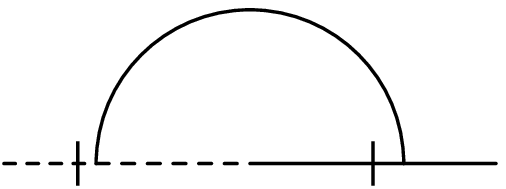}
   \quad &
   \includegraphics[width=2.3truecm]{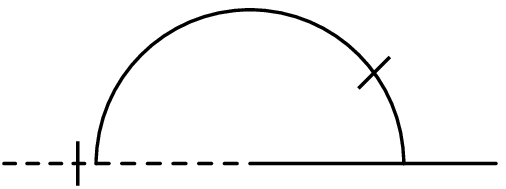}
   \quad &
    \end{tabular}
  \caption{One-loop approximation for the 1-irreducible function $\widetilde{\Gamma}(\varphi)$, entering into expression~\eqref{eq:expansion}}
  \label{Dia}
\end{figure}

The ultraviolet (UV) renormalizability can be revealed by the analysis
of the 1-irreducible Green functions~\cite{Vasiliev}. The corresponding generating functional can be written in the form
\begin{equation}
  \Gamma(\varphi) =  \S(\varphi) + \widetilde{\Gamma}(\varphi),
  \label{eq:expansion}
\end{equation}
where ${\S}(\varphi)$ stands for the action functional~\eqref{Fact}, \eqref{Dact}, or~(\ref{eq:action}) and
$\widetilde{\Gamma}(\varphi)$ is the sum of all the 1-irreducible diagrams~\cite{Vasiliev,Zinn}.
As it has been shown in~\cite{ANU97} and discussed in~\cite{AK14}, the reduced model~(\ref{eq:action})~--~(\ref{eq:correl2}) is
invariant with respect to the Galilean symmetry. This fact leads to the UV finiteness of the 
Green functions $v_i\partial_t v_i$ and $v'_i (v_j\partial_j)v_i$.
Therefore, there are only five Green functions which require renormalization:
\begin{equation}
\left\langle v'_iv_j \right\rangle_{\text{1-ir}}, \quad \left\langle \phi'\phi \right\rangle_{\text{1-ir}}, \quad  \left\langle v_i'\phi \right\rangle_{\text{1-ir}}, \quad  \left\langle \phi'v_i \right\rangle_{\text{1-ir}}, \quad  \text{and} \quad \left\langle  v_i'v_j' \right\rangle_{\text{1-ir}}.
\end{equation}
The one-loop diagrams which have to be calculated are presented in Fig.~\ref{Dia}.
Employing the dimensional regularization within minimal subtraction scheme (MS) one can calculate the renormalization constants. The UV 
divergences manifests themselves as pole terms in $y$ and $\eps=4-d$.
The poles can be eliminated by the multiplicative renormalization of the parameters $g_{10},g_{20},\nu_0,u_0,v_0,c_0$ and fields $\phi$ and $\phi'$:
\begin{align}
g_{10} &= g_1 \mu^y Z_{g_1},& \quad u_{0} &= u Z_{u}, &\quad \nu_0&=\nu Z_{\nu}, &\quad \phi &\to Z_{\phi}\phi,
\nonumber \\
g_{20} &= g_2 \mu^\varepsilon Z_{g_2}, &\quad v_{0} &= v Z_{v}, &\quad c_{0}&= c Z_{c}, &\quad \phi' &\to Z_{\phi'}\phi'.
\label{mult}
\end{align}
Here $\mu$ is the ``reference mass'' (additional free parameter
of the renormalized theory) in the MS renormalization scheme, 
the parameters $g_1,g_2,\nu,u,v$, and $c$ are renormalized analogs of the bare
parameters (without subscript ``$0$''), $Z_i$
are the renormalization constants, which depend only on the completely
dimensionless parameters $g_1,g_2,u,v,\alpha,d,y$ and $\varepsilon$. 
\newline

{\section{IR attractive fixed points} \label{sec:renorm}}
The large-scale behaviour with respect to the spatial and time variables is
governed by the IR attractive stable fixed points $g^*\equiv\{g_1^*,g_2^*,u^*,v^*\}$;
the asterisk refers to the coordinate of the fixed point (FP).
They are determined from the relations~\cite{Vasiliev,Zinn}:
\begin{align}
  &\beta_{g_1} (g^{*}) = \beta_{g_2} (g^{*})= \beta_{u}
  (g^{*}) = \beta_{v} (g^{*}) = 0,
  \label{eq:gen_beta}
\end{align}
where $\beta_x=\widetilde{D}_\mu x$ for any variable $x$, and the differential operator $\widetilde{D}_\mu $ denotes the operation
$\mu\partial_\mu$ for fixed bare parameters $\left\{g_{10}, g_{20}, u_0, v_0, \nu_0, \alpha_0, c_0\right\}$.
The eigenvalues of the matrix of the first derivatives 
$\Omega_{ij}=\left.\left({\partial \beta_i}/{\partial g_j}\right)\right|_{g=g^*}$, where $i,j\in\{g_1,g_2,u,v\}$,
determine whether the given FP is IR stable or not. Points with all positive eigenvalues are
candidates for macroscopic regimes and, in principle,
can be observed experimentally. 

\begin{figure}[b]
\center
\includegraphics[width=0.45\textwidth,clip]{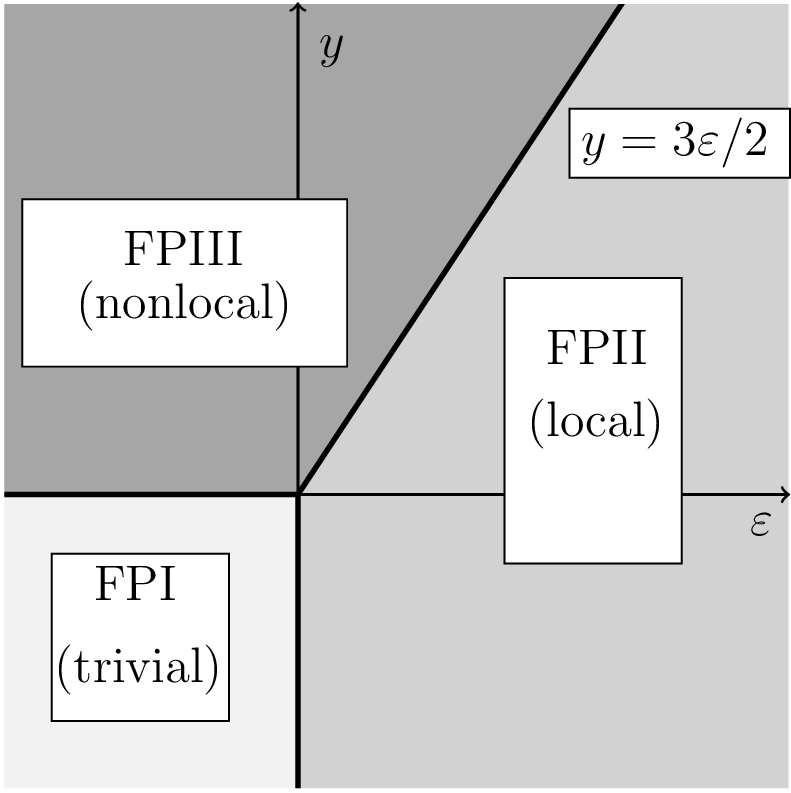}
\caption{Domains of IR stability of the fixed points in the model
(\protect\ref{eq:action})}
\label{fig:Plain}
\end{figure}

Explicit forms of the $\beta$-functions are
\begin{equation}
  \beta_{g_1} = g_1(-y - \gamma_{g_1}),\quad
  \beta_{g_2} = g_2(-\eps-\gamma_{g_2}),\quad
  \beta_u = -u\gamma_u,\quad
  \beta_v = -v\gamma_v,
  \label{eq:beta_functions}
\end{equation}
where $\gamma_i = \widetilde{D}_\mu \ln  Z_i$ are
the anomalous dimensions~\cite{Vasiliev}.
Unlike the simple case $d=3$, where only one nontrivial IR attractive fixed point exists, a direct analysis of the system of
equations~(\ref{eq:gen_beta}) reveals  existence of three IR stable fixed points: trivial one \fp{I} and two nontrivial, \fp{II} 
and \fp{III}; for details see~\cite{BigOne}. Corresponding domains of their IR stability are shown in Fig.~\ref{fig:Plain}. 
\newline
\newline
\newline
The fixed point \fp{I} with 
\begin{equation}
  g_1^* = 0, \quad g_2^* = 0
  \label{fp1}
\end{equation}
is stable if $y, \varepsilon<0$ with arbitrary $u^*$ and $v^*$; the fixed point \fp{II} with coordinates 
\begin{equation}
  g_1^* = 0, \quad g_2^* = \frac{8\eps}{3}, \quad
  u^* = 1, \quad
  v^* = 1
  \label{fp2}
\end{equation}
is stable if $\varepsilon>0$ and $y<3\varepsilon/2$; the fixed point \fp{III} with coordinates 
\begin{equation}
  g_1^* = \frac{16y(2y-3\eps)}{9(y(2+\alpha)-3\eps)}, 
  \quad g_2^* =\frac{16\alpha y^2}{9(y(2+\alpha)-3\eps)},
  \quad
  u^* = 1, \quad
  v^* = 1
  \label{fp3}
\end{equation}
is stable if $y>0$ and $\varepsilon<2y/3$.
The crossover between two nontrivial points occurs along the line $y=3\eps/2$, which is in accordance with~\cite{Ant04}.

This fact implies that depending on the values $y$ and $\varepsilon$ 
correlation functions of the model~(\ref{eq:action}) 
in the IR region exhibit different types of scaling behaviour. 
For the corresponding critical dimensions $\Delta[F]=d^{k}_{F}+ \Delta_{\omega}d^{\omega}_{F} + \gamma_{F}^{*}$, where $\gamma_{F}^{*}$ is the value
of the anomalous dimension $\gamma_{F}$ at the fixed point $g^*$ and $\Delta_{\omega}=2-\gamma_\nu^*$ is the critical
value of frequency, $d^{k}_{F}$ and $d^{\omega}_{F}$ are the frequency and momentum dimensions of $F$, we have 
$$
\Delta_{v}=1-\varepsilon/2, \quad \Delta_{v'}= d- \Delta_{v}, \quad
\Delta_{\omega}=2-\varepsilon/2, \quad \Delta_{m}=1;
$$
\begin{equation}
\Delta_{\phi}=d-\Delta_{\phi'}=2-5\varepsilon/4, \quad
\Delta_{c}= 1- 5\varepsilon/8
\label{Krit2}
\end{equation}
for the fixed point \fp{II};
$$
\Delta_{v}=1-y/3, \quad \Delta_{v'}= d- \Delta_{v}, \quad
\Delta_{\omega}=2-y/3, \quad \Delta_{m}=1;
$$
\begin{equation}
\Delta_{\phi}=d-\Delta_{\phi'}=2-5y/6, \quad
\Delta_{c}= 1- 5y/12
\label{Krit3}
\end{equation}
for the fixed point \fp{III}; more detailed discussion see in~\cite{AGKL1}.
\newline
{\section{Advection of a passive scalar field} \label{sec:adv}}

\begin{figure}[b]
\center
\includegraphics[width=0.20\textwidth,clip]{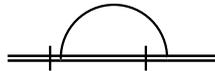}
\caption {One-loop approximation for the 1-irreducible function $\widetilde{\Gamma}(\varphi)$, responding to action functional~\eqref{Dact}}
\label{TT}
\end{figure}

Taking into our consideration a density field $\theta$ one should renormalize all diagrams, connected with the action
functional~\eqref{Dact}. From the dimensionality considerations together with symmetry reasons it follows that the
only function which requires the UV renormalization is the 1-irreducible function $\left\langle \theta'\theta\right\rangle_{\text{1-ir}}$. This 
situation is naturally reproduced by multiplicative
renormalization of the diffusion coefficient, $\kappa_{0}=\kappa Z_{\kappa}$. To calculate the renormalization constant $Z_\kappa$ in one-loop 
approximation one should analyze the diagram, shown in Fig.~\ref{TT}.

The function $\beta_{w} = \widetilde { D}_{\mu} w$ for the new
dimensionless parameter $w=\kappa/\nu$ has the form
\begin{equation}
\beta_{w} = - w\gamma_{w} = w (\gamma_{\nu}-\gamma_{\kappa}).
\label{betaw}
\end{equation}
Supplementing the system of equations~\eqref{eq:gen_beta} by the equation $\beta_{w}(g^*)=0$ one finds that the possible IR attractive
fixed points of the full problem~\eqref{Fact} are given by the expressions~\eqref{fp1}~-- ~\eqref{fp3} together with the condition $w^*=1$.

For the corresponding critical dimensions of the passive fields $\theta$ and $\theta'$ we have
\begin{equation}
\Delta_{\theta}= -1+\varepsilon/4, \quad \Delta_{\theta'}= d+1 -\varepsilon/4 \quad \text{for the fixed point \fp{II}};
\end{equation}
\begin{equation}
\Delta_{\theta}= -1+y/6, \quad \Delta_{\theta'}= d+1 -y/6  \quad \text{for the fixed point \fp{III}}.
\end{equation}

The measurable quantities are some correlation functions of composite operators.
A local composite operator is a monomial or polynomial constructed from the primary fields
$\Phi(x)$ and their finite-order derivatives at a single space-time point $x$. In the Green functions with such
objects, new UV divergences arise due to the coincidence of
the field arguments. They can be removed by the additional
renormalization procedure. 

\begin{figure}[t]
\center
\includegraphics[width=0.10\textwidth,clip]{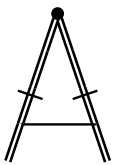}
\caption {One-loop approximation for the 1-irreducible function $\Gamma_{n}(x;\theta)$}
\label{OPD}
\end{figure}

In the following, the central role will be played by composite fields built solely from the basic fields~$\theta$: 
\begin{equation}
\label{comp}
F(x)=\theta^{n}(x).
\end{equation}
The renormalization constants $Z_n$ are determined by the requirement that the 1-irreducible functions $\Gamma_{n}(x;\theta)$
are UV finite in the renormalized theory, where $\Gamma_{n}(x;\theta)$ is the $\theta^{n}$ term of the
expansion in $\theta(x)$ of 
the generating functional of the
1-irreducible Green functions with one composite operator $F(x)$
and any number of fields $\theta$:
\begin{align}
\Gamma_{n}(x;\theta) =  \int {\mathrm d}x_{1} \cdots \int {\mathrm d}x_{n}
\theta(x_{1})\cdots\theta(x_{n})
\langle F(x) \theta(x_{1})\cdots\theta(x_{n})\rangle.
\label{Gamma1}
\end{align}
The operator diagram, needed to calculate the renormalization constants $Z_n$, is shown in Fig.~\ref{OPD}.
The critical dimensions $\Delta[\theta^{n}]$ of the operators $\theta^{n}$ are given by the expression 
\begin{equation}
\Delta[\theta^{n}]= n\Delta_{\theta} + \gamma_{n}^{*},
\label{KrOp}
\end{equation}
where the critical dimension $\Delta_{\theta}$ for the fixed point \fp{II} or \fp{III} is given by the expression~\eqref{Krit2} or~\eqref{Krit3}.
Finally, one obtains that
\begin{align}
\label{KrOp1}
\Delta\left[\theta^n\right]&=-n+\frac{n\varepsilon}{4}-\frac{n(n-1)}{3}\varepsilon \quad \text{for the fixed point \fp{II}};\\
\Delta\left[\theta^n\right]&=-n+\frac{ny}{6}-\frac{2n(n-1)}{3}\frac{\alpha y(y-\varepsilon)}{y(2+\alpha)-3\varepsilon} \quad \text{for the fixed point \fp{III}}.
\label{KrOp2}
\end{align}
Both the expressions~\eqref{KrOp1} and~\eqref{KrOp2} are supposed to contain higher-order corrections in $y$ and $\varepsilon$.
For the both cases \fp{II} and \fp{III} the dimensions are negative, i.e., ``dangerous'' in the sense of operator product expansion~\cite{Vasiliev,turbo},
and decrease as $n$ grows. 

From the RG representation together with dimensionality considerations it follows that at $\mu r \gg1$ the pair correlation
function of two composite operators $F(x)=\theta^n(x)$ takes the form
\begin{equation}
\langle \theta^{p}(t,{\bm x}_{1}) \theta^{k}(t,{\bm x}_{2}) \rangle \simeq
\mu^{-(p+k)} (\mu r)^{-\Delta_{p}-\Delta_{k}} \times h_{pk}[mr,\bar{c}(r)],
\label{Parr2}
\end{equation}
where $h_{pk}$ is unknown scaling function with completely dimensionless arguments, $\bar{c}(r)$ is invariant speed of sound. The inertial-convective range
$l\ll r\ll L$  corresponds to the additional condition $mr\ll1$. 
The behaviour of the functions $h_{pk}$ at $mr\to0$ can be
studied by means of the operator product expansion; see~\cite{Zinn,Vasiliev}.
According to the OPE, the equal-time product $F_{1}(x_1)F_{2}(x_2)$
of two renormalized operators for
${\bm x}\equiv ({\bm x_1} + {\bm x_2} )/2 = {\rm const}$ and
${\bm r}\equiv {\bm x_1} - {\bm x_2}\to 0$ has the representation
\begin{equation}
F_{1}(t,{\bm x}_{1}) F_{2}(t,{\bm x}_{2}) \simeq \sum_{F} C_{F}[mr, \bar{c}(r)] F(t,{\bm x}),
\label{OPE}
\end{equation}
where $C_{F}$ are the numerical coefficient
functions analytical in the variables $mr$ and $\bar{c}(r)$,
$F(t,{\bm x})$ are all possible renormalized local composite operators allowed by the symmetry.
Using the expansion~\eqref{OPE} one finally finds that if $l\ll r\ll L$ the asymptotic behaviour of the function~\eqref{Parr2} is given by the expression
\begin{equation}
\langle \theta^{p}(t,{\bm x}_{1}) \theta^{k}(t,{\bm x}_{2}) \rangle \simeq
\mu^{-(p+k)} (\mu r)^{-\Delta_{p}-\Delta_{k}} (mr)^{\Delta_{p+k}}.
\label{FinF}
\end{equation}
The leading term in the OPE is given by the operator from the same family with the index $p+k$. Moreover, the 
inequality $\Delta_{p}+\Delta_{k}>\Delta_{p+k}$ takes place. These two facts confirm the multifractal behaviour  of the scalar 
field in the model~\eqref{density1}; see~\cite{DL}.
\newline

{\section{Conclusion} \label{sec:consl}}

In this paper the model of turbulent advection of the passive scalar (density) field $\theta$ has been studied using the field theoretic approach. 
The velocity field is governed by the stochastic Navier-Stokes equation for a compressible fluid. 
In contrast to the previous works, the model is considered near the special space dimension $d=4$;
in this case the additional UV divergence is presented in the 1-irreducible Green function $\left\langle v'v'\right\rangle_{\text{1-ir}}$,
which significantly influence to RG analysis.
The crucial points of the Feynman diagrammatic technique and perturbative renormalization group
have been discussed. One loop approximation provides that 
for different meanings of the exponent $y$ and deviation from the dimension of ${\bm x}$ space $\eps=4-d$
the model possesses two nontrivial stable 
fixed points in the IR region~-- the local one (\fp{II}) and nonlocal one (\fp{III}).
This shows us, that the simple analysis around $d=3$, which indicates an existence of the only 
nontrivial fixed point, corresponding to nonlocal scaling regime~\cite{AK14}, is incomplete in this case.
Nevertheless, at most realistic values $y=4$ and $\varepsilon=1$ the new local regime is unstable, and well-known non-local
regime come true, therefore the main conclusions of the previous papers~\cite{ANU97,AK14} about existence and IR attraction of this fixed point remain true.

The inertial range $l\ll r\ll L$ behaviour of the correlation functions of the composite fields (operators), built solely from the impurity field $\theta$,
was studied by means of the OPE. The existence of the
of anomalous scaling, i.e., singular power-like dependence on the integral scale $L$, was established. 
In contrast to the case $d=3$, at $d=4$ the anomalous dimensions of the composite operators~\eqref{KrOp1} and~\eqref{KrOp2} do not grow with $\alpha$
without bound. This is a consequence of eliminating of poles in $\varepsilon$ near $d=4$, which leads to a significant improvement of situation near $d=3$.

It would be very interesting to go beyond the one-loop approximation and to discuss the existence, stability and the dependence 
on $\alpha$ of fixed points at the two-loop level, as well as to investigate a scalar admixture in case of tracer field, or passively advected vector fields.
Another important and very interesting problem is to develop the stochastic Navier-Stokes equation for a compressible fluid near $d=2$, which may show 
another types of IR behaviour. This work seems to be a difficult technical task and is left for the future.


\pagebreak
\begin{acknowledgement}

The authors are indebted to M.~Yu.~Nalimov, L.~Ts.~Adzhemyan, M.~Hnati\v{c}, J.~Honkonen, and V.~\v{S}kult\'ety for discussions.

The work was supported by VEGA grant No.~1/0222/13 of the Ministry
of Education, Science, Research and Sport of the Slovak Republic,
by the Saint Petersburg State University within the
research grant 11.38.185.2014, 
and by the Russian Foundation for Basic
Research within the Project 16-32-00086.
N.~M.~G. acknowledges the support from the Saint Petersburg Committee of Science and High School.
N.~M.~G. and M.~M.~K. were also supported by the Dmitry Zimin's ``Dynasty'' foundation.

\end{acknowledgement}



\begin{thebibliography}{}
\bibitem{Legacy} U.~Frisch, {\it Turbulence: The Legacy of A.~N.~Kolmogorov}
(Cambridge University Press, Cambridge, 1995)
\bibitem{Vasiliev} A.~N.~Vasil'ev, {\it The Field Theoretic  Renormalization Group in Critical Behavior Theory and Stochastic Dynamics} (Boca Raton, Chapman  Hall/CRC, 2004)
\bibitem{Zinn} J.~Zinn-Justin, \textit{Quantum Field Theory and Critical Phenomena} (4th edition, Oxford University Press, Oxford, 2002)
\bibitem{turbo} L.~Ts.~Adzhemyan, N.~V.~Antonov, A.~N.~Vasil'ev: \emph{The Field Theoretic Renormalization Group in Fully Developed Turbulence} (Gordon \& Breach, London, 1999)
\bibitem{Tauber} U.~T{\"a}uber, {\it Critical Dynamics: A Field Theory Approach to Equilibrium and Non-Equilibrium Scaling Behavior}  (Cambridge University Press, New York, 2014)
\bibitem{JphysA} N.V.~Antonov, J. Phys.~A: Math. Gen. {\bf 39}, 7825 (2006)
             
\bibitem{Kraich1} R.H.~Kraichnan, Phys. Fluids {\bf 11}, 945 (1968)

\bibitem{GK} K. Gaw\c{e}dzki and A.~Kupiainen, Phys. Rev. Lett. {\bf 75}, 3834 (1995); \\
D.~Bernard, K.~Gaw\c{e}dzki, and A.~Kupiainen, Phys. Rev.~E {\bf 54}, 2564  (1996); \\
M.~Chertkov and G.~Falkovich, Phys. Rev. Lett. {\bf 76}, 2706 (1996)

\bibitem{RG} L.~Ts.~Adzhemyan, N.~V.~Antonov, and A.~N.~Vasil'ev,
Phys. Rev.~E {\bf 58}, 1823 (1998)

\bibitem{JJ15} E.~Jur\v{c}i\v{s}inova, M.~Jur\v{c}i\v{s}in, Phys. Rev.~E {\bf 91}, 063009 (2015)


\bibitem{V96-mod} N.~V.~Antonov,   A.~Lanotte, and  A.~Mazzino, Phys. Rev.~E {\bf 61},  6586 (2000); \\
N.~V.~Antonov, N.~M.~Gulitskiy, Theor. Math. Phys., {\bf 176}(1), 851 (2013)


\bibitem{amodel}  
H.~Arponen, Phys. Rev.~E, {\bf 79},  056303 (2009)



\bibitem{AG12}
N.~V.~Antonov, N.~M.~Gulitskiy, Lecture Notes in Comp. Science,
{\bf 7125/2012},  128 (2012); \\
N.~V.~Antonov, N.~M.~Gulitskiy, Phys. Rev.~E {\bf 85},  065301(R) (2012); \\
N.~V.~Antonov, N.~M.~Gulitskiy, Phys. Rev.~E {\bf 87},  039902(E) (2013)


\bibitem{J13}
E.~Jur\v{c}i\v{s}inova, M.~Jur\v{c}i\v{s}in, J. Phys.~A: Math. Theor., {\bf 45},  485501 (2012); \\ 
E.~Jur\v{c}i\v{s}inova, M.~Jur\v{c}i\v{s}in, Phys. Rev.~E {\bf 88}, 011004 (2013)



\bibitem{Uni3} E.~Jur\v{c}i\v{s}inova and M.~Jur\v{c}i\v{s}in, Phys. Rev.~E {\bf 77}, 016306 (2008); \\
E.~Jur\v{c}i\v{s}inova, M.~Jur\v{c}i\v{s}in and R.~Remecky, Phys. Rev. E {\bf 80}, 046302 (2009)

\bibitem{VectorN}   
N.~V.~Antonov and N.~M.~Gulitskiy, Phys. Rev.~E {\bf 91},  013002 (2015);\\
N.~V.~Antonov and N.~M.~Gulitskiy, Phys. Rev.~E {\bf 92},  043018 (2015);\\
N.~V.~Antonov and N.~M.~Gulitskiy, AIP Conf. Proc. {\bf 1701}, 100006 (2016); \\
N.~V.~Antonov and N.~M.~Gulitskiy, EPJ Web of Conf. {\bf 108}, 02008 (2016)

\bibitem{NSpass} L.~Ts.~Adzhemyan, N.~V. Antonov, J.~Honkonen, and T.~L.~Kim,  Phys. Rev.~E {\bf 71}, 016303 (2005)

\bibitem{Ant04} N.~V.~Antonov, Phys. Rev. Lett. {\bf 92}, 161101 (2004)

\bibitem{AK14} N.~V.~Antonov and M.~M.~Kostenko, Phys. Rev.~E {\bf 90}, 063016 (2014)

\bibitem{AK15} N.~V.~Antonov and M.~M.~Kostenko, Phys. Rev.~E {\bf 92}, 053013 (2015)   

\bibitem{AGM} N.~V.~Antonov, N.~M.~Gulitskiy, and A.~V.~Malyshev, EPJ Web of Conf. {\bf 126}, 04019 (2016)

\bibitem{JJR16} E.~Jur\v{c}i\v{s}inova, M.~Jur\v{c}i\v{s}in, R.~Remecky, Phys. Rev.~E {\bf 93}, 033106 (2016)


\bibitem{VM} M.~Vergassola and A.~Mazzino, Phys. Rev. Lett. {\bf 79}, 1849 (1997)


\bibitem{tracer2} A.~Celani, A.~Lanotte, and A.~Mazzino,
Phys. Rev.~E {\bf 60} R1138 (1999)

\bibitem{tracer3} M.~Chertkov, I.~Kolokolov, and M.~Vergassola,
Phys.~Rev.~E. {\bf 56}, 5483 (1997) 






\bibitem{AK2}  M.~Hnatich, E.~Jur\v{c}i\v{s}inova, M.~Jur\v{c}i\v{s}in,
and M. Repa\v{s}an, J. Phys. A: Math. Gen. {\bf 39}, 8007 (2006)

\bibitem{LM} V.~S.~L'vov and A.~V.~Mikhailov, Preprint No. 54, Inst. Avtomat. Electron., Novosibirsk (1977)
\bibitem{St} I.~Staroselsky, V.~Yakhot, S.~Kida, and S.~A.~Orszag, Phys. Rev. Lett., {\bf 65}, 171 (1990)
\bibitem{MTY} S.~S.~Moiseev, A.~V.~Tur, and V.~V.~Yanovskii, Sov. Phys. JETP {\bf 44}, 556 (1976)

\bibitem{ANU97} N.~V.~Antonov, M.~Yu.~Nalimov and A.~A.~Udalov, Theor. Math. Phys. {\bf 110}, 305 (1997)
\bibitem{HN96} J.~Honkonen and M.~Yu.~Nalimov, Z. Phys. B {\bf 99}, 297 (1996)
\bibitem{LL} L.~D.~Landau and E.~M.~Lifshitz, {\it Fluid Mechanics}
(Pergamon Press, Oxford)
\bibitem{VN96} D.~Yu.~Volchenckov and M.~Yu.~Nalimov, Theor. Math. Phys. {\bf 106}, 307 (1996)

\bibitem{AHKV03} 
 L.~Ts.~Adzhemyan, J.~Honkonen, M.~V.~Kompaniets et al., Phys. Rev. E {\bf 71}, 036305 (2005)
\bibitem{AHH10} L.~Ts.~Adzhemyan, M.~Hnatich and J.~Honkonen, Eur. Phys. J B {\bf 73}, 275 (2010)

\bibitem{BigOne} N.~V.~Antonov, N.~M.~Gulitskiy, M.~M.~Kostenko, and T. Lu\v{c}ivjansk\'y, in preparation; see arXiv:1611.00327

\bibitem{AGKL1} N.~V.~Antonov, N.~M.~Gulitskiy, M.~M.~Kostenko, and T. Lu\v{c}ivjansk\'y, EPJ Web of Conf. {\bf 125}, 05006 (2016)
\bibitem{DL}
B.~Duplantier, A.~Ludwig, Phys. Rev. Lett.  {\bf 66},  247 (1991);\\
G.~L.~Eyink, Phys. Lett.~A {\bf 172}, 355 (1993)





\end{thebibliography}
\end{document}